\documentstyle[epsf,rawfonts]{EuroPhys} 
\input EuroMacr
\begin{document}
\euro{}{}{}{}
\Date{}
\shorttitle{K.K.NG \etal  }
\title{The role of spin-orbit coupling for the superconducting state in Sr$_2$RuO$_4$}
\author{Kwai Kong Ng and Manfred Sigrist}
\institute{Yukawa Institute for Theoretical Physics, Kyoto University,
Kyoto 606-8502, Japan}
\rec{}{}
\pacs{\Pacs{71}{70.Ej}{Spin-orbit coupling, Zeeman and Stark
splitting, Jahn-Teller effect }  
\Pacs{74}{25.Dw}{Superconductivity phase diagrams}} 
\maketitle
\begin{abstract} 
The odd-parity spin-triplet Cooper pairing states show a six-fold
degeneracy for a quasi-two-dimensional electron system. In this
article we show how spin-orbit coupling can lift this degeneracy
for Sr$_2$RuO$_4$ taking the band structure based on
the three relevant $ t_{2g} $-orbitals of the 
Ru-ions into account. The stabilized state depends on the relative
strength of the pairing interaction. We show that under reasonable
assumptions the chiral pairing state $ {\bf d} ({\bf k} ) = \hat{{\bf
z}} (k_x \pm i k_y) $ is favored against the others.

\end{abstract}

The search for new superconducting transition metal oxide systems has
lead a few years ago to the discovery of Sr$_2$RuO$_4$
\cite{MAENO1}. Despite the low transition temperature, ($ T_{cmax} \approx
1.5 K$ ) which is, in addition, very sensitive to disorder \cite{MACKENZIE}, 
the unusual properties of the superconductor have attracted much
interest. During the last year it has been experimentally established
that this compound is a spin-triplet superconductor 
as initially suggested on theoretical basis \cite{RICE,BASK}. The 
superconducting state is characterized by the violation of 
time-reversal symmetry \cite{LUKE} and equal spin pairing within the
basal plane of the tetragonal crystal lattice \cite{ISHIDA}. The
single candidate consistent with all presently known experimental data 
is described by the gap function of the symmetry $ {\bf d} ({\bf k}) =
\hat{{\bf z}} ( k_x \pm i k_y ) $ in the standard vector notation
\cite{HERAKLION}. This state is two-fold degenerate and has chiral
symmetry, i.e. the Cooper pairs possess an orbital angular momentum. 
There are various possible other candidates for spin-triplet pairing
which might be rather close in energy as we will show below. The
question arises what is the microscopic reason to favor the 
chiral phase compared to the others. It is the aim of this paper to
show that the chiral state is a natural consequence of the electronic
structure of Sr$_2$RuO$_4$ and the presence of spin-orbit coupling.

Sr$_2$RuO$_4$ is like several of the high-temperature superconductors
a layered perovskite system of stacking RuO$_2$-planes with a single
layer per unit cell \cite{MAENO1}. Like Cu in 
the high-temperature superconductors Ru forms a square lattice. While 
a single-band description seems to be adequate for the CuO$_2$-system
the case of RuO$_2$ requires at least three bands originating from the
three 4d-$t_{2g} $-orbitals of Ru occupied by four electrons on the
average. All bands cross the Fermi energy and give rise to three 
Fermi surfaces, two electron-like and one hole-like, all of nearly
cylindrical shape owing to the weak dispersion along the $c$-axis 
\cite{OGUCHI,SINGH,DHVA}. The low-temperature normal state of
Sr$_2$RuO$_4$ is that of a strongly correlated, nearly two-dimensional
Fermi liquid. In this sense this system resembles to some extent the
nature of the Fermi liquid $^3$He \cite{RICE}.

Starting from this assumption we show that the ordinary weak coupling
theory for spin-triplet pairing yields a six-fold degenerate
superconducting state if the spin space is completely rotation
symmetric. This degeneracy is lifted through spin-orbit 
coupling as suggested in Ref.(\cite{RICE,HERAKLION}). In this letter
we describe the effect of spin-orbit coupling based on the
orbital origin of the three electron bands. For this purpose we
introduce a simple model taking the basic structure of the three
orbitals into account and leading to spin-triplet superconductivity.
The three $ t_{2g} $-orbitals, $ d_{yz} , d_{zx} , d_{xy} $, will be
labeled as $ 1,2,3 $, respectively. 
The kinetic energy is then given by

\begin{equation}
{\cal H}_{kin} = \frac{1}{N} \sum_{{\bf k},s} C^{\dag}_{{\bf k}s} \left
( \begin{array}{ccc}
\varepsilon_{{\bf k}1} & g_{\bf k} & 0 \\
g_{\bf k} & \varepsilon_{{\bf k}2} & 0 \\
0 & 0 & \varepsilon_{{\bf k}3} \end{array} \right) C_{{\bf k}s}
\label{kinen}
\end{equation}
with $ C^{\dag}_{{\bf k}s} = (c^{\dag}_{{\bf k}1s} c^{\dag}_{{\bf k}2s}
c^{\dag}_{{\bf k}3s}) $, $ (\varepsilon_{{\bf k}1}, \varepsilon_{{\bf
k}2}) = - 2 t_1 ( \cos k_y , \cos k_x  )- \mu $, $
\varepsilon_{{\bf k}3} = -2 t_2 (\cos k_x + \cos k_y ) - 4 t_3 \cos k_x 
\cos k_y  - \mu' $ and $ g_{{\bf k}} = - 4 t_4 \sin k_x \sin k_y $ ($N$: 
number of lattices sites). 
The parameters will be later chosen to fit the band structure as close as
possible \cite{SINGH,XIANG}. Through the hybridization 
of the orbital $ 1 $ and $ 2 $ the Hamiltonian leads to the bands $ \alpha $ 
and $ \beta $ corresponding to the hole-like and
electron-like Fermi surface, respectively.  
The band belonging to orbital $ 3 $ is called $ \gamma $ in
accordance with the band structure literature \cite{OGUCHI,SINGH,XIANG}. 

We now turn to the discussion of the pairing interaction yielding
spin-triplet pairing. The onsite
interaction is definitely repulsive suppressing pairing in the
standard $s$-wave channel. Therefore we focus on nearest-neighbor-site
interactions. Furthermore, the zero-momentum pairing requires that 
two paired electrons are on the same Fermi surface. Interband pairs
have, in general, a finite net momentum. 
As a possible candidate for pairing ferromagnetic
spin fluctuations were discussed \cite{HERAKLION,SF1}. Experimental studies 
show, however, that the ferromagnetic component of the spin
fluctuations are rather weak \cite{BRADEN} and, in 
additions, if they are enhanced, e.g. by pressure, superconductivity is
actually suppressed \cite{PRESSURE}. Alternatively, Hund's rule
coupling was mentioned as another cause for parallel spin pairing
\cite{RICE,BASK}. Standard Hund's rule coupling on the Ru-ion,
however, yields an onsite interaction and would not be available for
spin-triplet pairing. 
There is, however, a spin-dependent interaction mediated via
orthogonal oxygen orbitals between different nearest-neighbor
orbitals. For example, on a bond between site 1 and 2  along the 
$ x $-axis spins on the Ru-$ d_{xy} $- and $ d_{xz} $-orbital on
different sites interact ferromagnetically as they hybridize with orthogonal
$p $-orbitals on the intermediate oxygens. This is a variant of the
Goodenough-Kanamori mechanism for ferromagnetic exchange based on
Hund's rule effect on oxygen in transition metal oxides. It is easy
to see that the $ d_{xy} $-orbital couples in this way to the
$ d_{zx} $- and $ d_{yz} $-orbital of the Ru-ions on 
neighboring sites along the $ x $- and $ y $-direction,

\begin{equation}  
H_{int} = J' \sum_i [ {\bf S}_{i3} \cdot {\bf S}_{i+\hat{\bf x},2} +
{\bf S}_{i+\hat{\bf x},3} \cdot {\bf S}_{i2} +  
{\bf S}_{i3} \cdot {\bf S}_{i+\hat{\bf y},1} +
{\bf S}_{i+\hat{\bf y},3} \cdot {\bf S}_{i1} ]
\end{equation}
This is an interorbital interaction and by itself for pairing not so
effective. However, the local spin polarization generated by this
interaction can be transfered via onsite Hund's rule coupling to the
other orbitals and lead to an effective spin dependent nearest
neighbor intraorbital pairing interaction which has the form

\begin{equation} 
H_{p}= \sum_i [ J_1 \{{\bf S}_{i1} \cdot {\bf S}_{i+\hat{\bf x}1} + 
{\bf S}_{i2} \cdot {\bf S}_{i+\hat{\bf y}2} \}
+ J_2 \{{\bf S}_{i1} \cdot {\bf S}_{i+\hat{\bf y}1} +
{\bf S}_{i2} \cdot {\bf S}_{i+\hat{\bf x}2} \}  
+ J_3 \{ {\bf S}_{i3} \cdot {\bf S}_{i+\hat{\bf x}3} + 
{\bf S}_{i3} \cdot {\bf S}_{i+\hat{\bf y}3} \} ]
\label{exchange}
\end{equation}

where $J_1$ is ignored as it has no contribution in our model. 
The effective coupling strengths depend on the onsite polarizability
of the spins in the different orbitals.

\begin{equation} 
J_2 = -\frac{J' J_H}{N} \sum_{{\bf q}} \chi_{33} ({\bf q}) \qquad {\rm
and } \qquad 
J_3 = - \frac{J' J_H}{N} \sum_{{\bf q}} [ \chi_{11} ({\bf q}) +
\chi_{12} ({\bf q}) ]
\label{J2J3}
\end{equation}
where $ \chi_{\mu \nu} ({\bf q}) = -\frac{i}{N}\int_0^{\infty} dt \langle [{\bf
S}_{\mu {\bf q}}(t), {\bf S}_{\nu -{\bf q}}(0)] \rangle $ is the
static spin susceptibility (ignoring retardation effects) of the orbital $ \mu
$ with the field acting on the orbital $\nu$.

Besides this spin dependent contribution we introduce  also a spin
independent part of the attractive intraorbital interaction of unknown
origin. This can be done by replacing $ {\bf S}_{i \mu} \cdot {\bf
S}_{j \mu} $ by $ {\bf S}_{i \mu} \cdot {\bf
S}_{j \mu} + v_\mu n_{i \mu } n_{j \mu}/4 $. 
Decomposed into spin-singlet and spin-triplet channel, this does not change
the structure of the interaction. For simplicity, we will take $
v_1=v_2=v_3 $ for our calculation. 
To lift the degeneracy completely we have also to include the pair
scattering between different Fermi surfaces. This interaction is
mediated by Coulomb interaction \cite{ODS}. For unconventional Cooper
pairs the onsite scattering is ineffective. Thus we include the
scattering via nearest-neighbor interaction also here,

\begin{equation}
H_{ib} = \sum_{i, \mu, \mu'} \sum_{s,s',\hat{\bf r}} g
c^{\dag}_{i \mu s} c^{\dag}_{i + \hat{\bf r},\mu s'}  c_{i + \hat{\bf
r}, \mu' s'} c_{i \mu' s}.
\label{interband}
\end{equation}
for $ \mu \neq \mu' $. Including all terms and changing to momentum
space we obtain in a pairing interaction of the form,

\begin{equation}
H_{pair} = \frac{1}{N} {\sum}^{'} V^r_{\mu \mu',s_1, ... s_4}
\cos (k_r - k_r') c^{\dag}_{{\bf k} \mu s_1} 
c^{\dag}_{-{\bf k} \mu s_2} c_{- {\bf k}' \mu' s_3} c_{{\bf k}' \mu'
s_4} 
\end{equation}
where $ \sum' $ indicates that we sum over all repeated indices ($ r = 
x,y $) and the coefficients $ V^r_{\mu \mu',s_1, ... s_4} $ are easily 
obtained from the above interactions.  The decomposition of $
\cos(k-k') $ into $ \cos k \cos k' + \sin k \sin k' $ yields the wave
functions of the possible states: ``extended $s$-wave'' ($ \cos k_x +
\cos k_y $), ``$ d_{x^2-y^2} $-wave'' ($ \cos k_x - \cos k_y $) for the
spin-singlet channel and the $ p $-wave'' spin-triplet states ($ \sin
k_x , \sin  k_y $) which are degenerate by symmetry
\cite{MIYAKE}. While both spin-singlet 
states have, in general, zero-nodes in the gap, it is possible to
combine the two degenerate $ p $-wave components to form a nodeless
gap. In this sense the $ p $-wave channel is expected to be favored, 
if the spin dependent part of the interaction does not suppress the triplet spin
configuration.

The degeneracy of the two Cooper pair channels with the orbital wave
functions $ \sin k_x $ and $ \sin k_y $ is independent of the spin
wave function. Thus, we may choose the spin configuration to optimize
the condensation energy. The gap function is given by

\begin{equation}
\hat{\Delta}_\mu ({\bf k})  = i \hat{\mbox{\boldmath
$ \sigma  $}} \cdot {\bf d}_\mu ({\bf k}) \hat{\sigma}^y
 = i \sum_{\nu=x,y,z} \hat{\sigma}^{\nu} 
\hat{\sigma}^y \hat{\mbox{\boldmath $ \nu  $}} (a_{\nu \mu} \sin k_x 
+ b_{\nu \mu} \sin k_y) 
\end{equation}
where $ \hat{\mbox{\boldmath $ \nu  $}} $ is the unit vector in $ \nu
$-direction. The quasiparticle energy gap in each band is given by $
\sqrt{|{\bf d}_\mu ({\bf k})|^2} $. States with the same gap have within the
weak coupling theory the same condensation energy and are consequently 
degenerate. The combinations given in Table I have all the same gap
$ \sqrt{\sin^2k_x + \sin^2k_y} $ so that we find six degenerate
pairing states \cite{RICE,HERAKLION,MIYAKE}. This is a special feature of a
two-dimensional Fermi liquid. In three dimensions one finds 
only one stable state, the Balian-Werthamer state, $ {\bf d} ({\bf
k}) = \hat{{\bf x}} \sin k_x + \hat{{\bf y}} \sin k_y + \hat{{\bf z}}
\sin k_z $. Note, that this result is general and does not depend on
the details of the Hamiltonian as long as it favors $p$-wave pairing. 
It was shown that going beyond weak coupling by including the
renormalization of the spin fluctuation spectrum the state $ {\bf d}
({\bf k}) = \hat{{\bf
z}}(\sin k_x \pm i \sin k_y ) $ would be favored analogous to the spin
fluctuation feedback mechanism stabilizing the A-phase of superfluid
$^3$He \cite{HERAKLION}. This is a secondary effect and does not 
affect the degeneracy of $ T_c $, but is connected with the
condensation energy at low temperature.

\begin{table}
\begin{center}
\caption{Degenerate stable spin-triplet pairing states listed
according to the irreducible representations of the tetragonal point
group $ D_{4h} $ assuming simultaneous rotations of spin and orbital
part. } 
\vspace{0.5cm}
\begin{tabular}{|c|c|cc|} \hline
{$ \Gamma $}  & {$ {\bf d}_\mu ({\bf k})$} & $P$ & $Q$ \\ \hline 
{$ A_{1u} $}  & {$ \hat{{\bf x}}\sin k_x + \hat{{\bf y}} \sin k_y $} & 
$+1$ & $+1$ \\ 
{$ A_{2u} $}  & {$ \hat{{\bf x}}\sin k_y - \hat{{\bf y}} \sin k_x $} & 
$+1$ & $-1$ \\ 
{$ B_{1u} $}  & {$ \hat{{\bf x}}\sin k_x - \hat{{\bf y}} \sin k_y $} & 
$-1$ & $+1$ \\ 
{$ B_{2u} $}  & {$ \hat{{\bf x}}\sin k_y + \hat{{\bf y}} \sin k_x $} & 
$-1$ & $-1$ \\ 
{$ E_u    $}  & {$ \hat{{\bf z}}(\sin k_x \pm i \sin k_y ) $} & $\pm1$
& - \\
\hline\end{tabular}
\end{center}
\vspace{-0.5cm}
\end{table}
The degeneracy of $ T_c $ can be lifted by removal of spin rotation
symmetry, i.e. by taking spin-orbit coupling into account.
The spin-orbit coupling is connected with the heaviest ion - Ru in our
case - where it acts as an onsite term in the Hamiltonian, 
$ {\cal H}_{so} = \lambda \sum_i {\bf  L}_i \cdot {\bf S}_i $. The angular
momentum $ {\bf L}_i $ operates on the three $ t_{2g}
$-orbitals on the site $ i $. If restrict ourselves to these three orbitals,
ignoring the $ e_g $-orbitals, we can write the spin-orbit Hamiltonian
as

\begin{equation}
{\cal H}_{so} = i \frac{\lambda}{2} \sum_{l,m,n} \epsilon_{lmn} \sum_{{\bf
k},s,s'} c^{\dag}_{{\bf k} l, s} c_{{\bf k} m s'}
\sigma^n_{ss'} 
\label{so}
\end{equation}
where $ \epsilon_{lmn} $ is the completely antisymmetric tensor and $
\lambda $ is a coupling constant such that the $ t_{2g} $-states
behave like an $ \ell=1 $ angular momentum representation (we use for $
l,m,n $ either $ \{ x,y,z \} $ or equivalently $ \{1,2,3 \} $).
Together with Eq.(\ref{kinen}) the
spin-orbit coupling leads to new
quasiparticles which are now labeled by pseudo-orbital and pseudo-spin indices
connected with the original ones by 
the unitary transformation defined for each wave vector $ {\bf k}
$,

\begin{equation}
( a_{{\bf k} 1 s}, a_{{\bf k} 2 s}, a_{{\bf
k} 3 s} )^{\dag} = ( c_{{\bf k} 1 s} , c_{{\bf k} 2 s} , c_{{\bf
k} 3 -s} )^{\dag} \hat{U}_{{\bf k} s}
\label{unitary}
\end{equation}
where the spin-orbit coupling mixes the up spins of the orbital 1 and
2 with down spin of orbital 3 and vice versa. We find that the
matrix elements of $ \hat{U}_{{\bf k} s} $ of opposite spin are
related as, $ u_{{\bf k} \mu \nu s} = u^*_{{\bf k} \mu \nu, -s} \quad {\rm for}
\quad \mu = 1, 2 $ and $ u_{{\bf k} 3 \nu s} = -u^*_{{\bf k} 3 \nu, -s} \quad {\rm for} \quad
\mu=3 $. 
Note, that the inclusion of spin-orbit coupling affects the band
structure so that the chemical potential of the orbitals have to be
adjusted for different values of  $ \lambda $ to produce the proper
Fermi surface shape.

Cooper pairing has now to be defined and analyzed concerning
the symmetry based on these new quasiparticles. We proceed, therefore, 
by rewriting the original pairing interaction $H_{pair}$ in terms of
these quasiparticles,

\begin{equation}
H_{pair} = \frac{1}{2 N}  {\sum}^{'} {\cal V}^{\nu
\nu'}_{s_1 s_2 s_3 s_4} ( {\bf k}, {\bf k}') 
 a^{\dag}_{{\bf k} \nu s_1} 
 a^{\dag}_{-{\bf k} \nu s_2} a_{-{\bf k}' \nu' s_3} a_{{\bf k}' \nu'
 s_4}   
\end{equation}
where
\begin{eqnarray}
{\cal V}^{\nu
\nu'}_{s_1 s_2 s_3 s_4} ( {\bf k}, {\bf k}') & = &\left( K^{m m'}_{s_1 s_2 s_3
 s_4} {\cal 
 U}^{P*}_{ {\bf k} m \nu s_1 s_2} {\cal U}^{P}_{ {\bf k'} m' \nu' s_3 s_4
} + \tilde{K}^{m m'}_{s_1 s_2 s_3 s_4} {\cal U}^{P*}_{{\bf k} m
 \nu s_1 s_2 }{\cal  U}^{-P}_{{\bf k'} m' \nu' s_3 s_4 } \right) \nonumber \\
K^{m m'}_{s_1 s_2 s_3 s_4} &=& \tilde{J}_{m}\delta_{m m'}
{\bf \sigma}_{s_1 s_4} {\bf \sigma}_{s_2 s_3} + \tilde{g}_{m m'}
\delta_{s_1 s_4} \delta_{s_2 s_3} + \tilde{g}'_{m m'}\delta_{s_1,
-s_4} \delta_{s_1, -s_2} \delta_{s_3,-s_4}  \nonumber \\
\tilde{K}^{m m'}_{s_1 s_2 s_3 s_4} & = & \tilde{g}'_{m m'} \delta_{s_1,
-s_4} \delta_{s_1, s_2} \delta_{s_3,s_4} 
\end{eqnarray}
Here we introduced the abbreviation $ \tilde{J}_{m} =
(J_{m}/4)(1+v) $. $\tilde{g}$ and $\tilde{g}'$ are symmetric with
non-vanishing elements $\tilde{g}_{1 2}=\tilde{g}'_{1 3}=\tilde{g}'_{2 3}=g$. 
The functions ${\cal U}^P$, $P=\pm 1$, are used to express the pairing
states and shall be constructed according to the
symmetries listed in Table I where $ P $ for each state
is given, 

\begin{eqnarray}
{\cal U}^P_{{\bf k} 1 \nu s_1 s_2} &=&  u_{{\bf k} 1 \nu s_1} u_{{\bf
k} 1 \nu s_2 } \sin k_x + i P u_{ {\bf k} 2 \nu s_1} u_{{\bf
k} 2 \nu s_2 } \sin k_y  \nonumber \\ 
{\cal U}^P_{ {\bf k} 2 \nu s_1 s_2} &=& u_{ {\bf k} 2 \nu s_1} u_{{\bf
k} 2 \nu s_2 } \sin k_x + i P u_{ {\bf k} 1 \nu s_1} u_{{\bf
k} 1 \nu s_2 } \sin k_y  \nonumber \\ 
{\cal U}^P_{ {\bf k} 3 \nu s_1 s_2} &=& u_{{\bf k} 3 \nu s_1 } u_{{\bf
k} 3 \nu s_2 }( \sin k_x - i s_1 s_2 P \sin k_y)
\end{eqnarray}
A $90^o$ rotation acting on the orbital part only (i.e. {\bf
k}) leads to ${\cal U}^P_{{\bf k} m \nu s_1 s_2} \rightarrow - i P {\cal U}^P_{{\bf k} 
m \nu s_1 s_2}$ following Table I. Note, that in $H_{pair}$ different
signs of $P$ are coupled through the pair scattering between orbital $3$ and
orbital $1$ or $2$ only. These interband pair scatterings are 
responsible for the lifting of degeneracy of the $A_{1,2 u}$ and
$B_{1,2 u}$ pairing states. 

The standard BCS-type mean-field approach leads to the definition of
the quasiparticle gap,

\begin{equation}
\Delta_{\nu s_2 s_1}({\bf k})= \frac{1}{4N} {\sum}^{'} {\cal V}^{\nu
\nu'}_{s_1,s_2,s_3,s_4} ( {\bf k}, {\bf k}') f_{\nu' s_3 s_4}({\bf k}')
\end{equation}
with $f_{\nu' s_3 s_4}({\bf k}')=\langle a_{-{\bf k}' \nu' s_3} a_{{\bf
k}' \nu' s_4} \rangle$ which is readily calculated after diagonalizing 
the mean-field Hamiltonian. After some
algebra, we formulate the self-consistent gap equation for pairing in the 
odd-parity (pseudo spin-triplet) channel for the $ {\bf d} $-vector,

\begin{eqnarray}
d^x_{\nu}({\bf k})&=& \frac{1}{2N} {\sum}^{'}\left
(  \tilde{J}_{m}
\delta_{m m'} + \tilde{g}_{m m'} - Q \tilde{g}'_{m m'}) \right) \left( Q {\cal
U}^{-P*}_{{\bf k} m \nu \downarrow \downarrow} + {\cal U}^{P*}_{{\bf k} 
m \nu \uparrow \uparrow} \right) D^{PQ}_{m'} \nonumber \\
d^y_{\nu}({\bf k})&=&\frac{i}{2N} {\sum}^{'}\left(  \tilde{J}_{m}
\delta_{m m'} + \tilde{g}_{m m'} - Q \tilde{g}'_{m m'} \right) \left( - Q {\cal
U}^{-P*}_{{\bf k} m \nu \downarrow \downarrow} + {\cal U}^{P*}_{
{\bf k} m \nu \uparrow \uparrow} \right) D^{PQ}_{m'} \nonumber \\
d^z_{\nu}({\bf k})&=&-\frac{1}{N} {\sum}^{'}\left(  \tilde{J}_{m}
\delta_{m m'} + \tilde{g}_{m m'} + \tilde{g}'_{m m'} \right) {\cal
U}^{P*}_{{\bf k} m \nu \uparrow \downarrow} D^P_{m'} 
\end{eqnarray}
with $D^{PQ}_{m'}$ and $ D^P_{m'}$ are, respectively, the
eigenvectors of 
 
\begin{eqnarray}
D^{PQ}_{m '} &=& - \frac{1}{N} {\sum}^{'}\left( \tilde{J}_{m}
\delta_{m m''} + \tilde{g}_{m m''} - Q \tilde{g}'_{m m'} \right)
 {\cal U}^P_{{\bf k} m' \nu' \uparrow \uparrow} {\cal
U}^{P*}_{{\bf k} m \nu' \uparrow \uparrow} F_{\nu' {\bf k}} 
D^{PQ}_{m''} \nonumber \\
D^P_{m'} &=&  - \frac{1}{N} {\sum}^{'}\left
( \tilde{J}_{m} \delta_{m m''} + \tilde{g}_{m m''} + \tilde{g}'_{m m'} \right)
{\cal U}^P_{{\bf k} m' \nu' \uparrow \downarrow} {\cal
U}^{P*}_{{\bf k} m \nu' \uparrow \downarrow} F_{\nu' {\bf k}}  D^P_{m''}
\end{eqnarray}
where $ F_{\nu, {\bf k}} = \tanh (\beta E_{\nu' {\bf k}}/2)/ 2
E_{\nu'{\bf k}} $ and $ P, Q = \pm 1$ corresponding to Table I. These
six eigen-equations (4 for 
$D^{PQ}_{m'}$ and 2 for $ D^P_{m'}$) has solutions with the
symmetries corresponding to the pairing states as in Table I. Note that
$ D^P_{m'}$ is independent of $Q$ and is the same for both
$P=\pm 1$ (because ${\cal U}^{-P}_{{\bf k} m \nu \uparrow \downarrow}={\cal
U}^{P*}_{{\bf k} m \nu \uparrow \downarrow}$ ), which is, therefore, always
doubly degenerate. If $ D^P_{m'}$ gives the highest
transition temperature $T_c$ after solving the eigen-equations
self-consistently, then all $D^{PQ}_{m'}$ has only trivial
solutions at $T_c$. Hence the pairing state has only $d^z$ component
and is doubly degenerate, which is consistent to the chiral state. On 
the other hand,  there will be no $d^z$ component in the pairing
state when one of the $D^{PQ}_{m'}$ gives the highest $T_c$. 
\begin{figure}[htbp]
\vspace{0.5cm}
\begin{center}
\epsfxsize=0.5\hsize
\epsffile{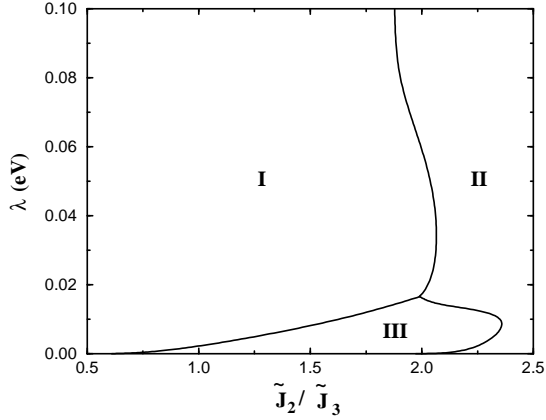}
\end{center}
\caption{Phase diagram of stable states, with phase I represents 
$\hat{{\bf z}} ( k_x \pm i k_y )$, phase II $\hat{\bf x}k_y - \hat{\bf
y}k_x$ and phase III $\hat{\bf x}k_x + \hat{\bf y}k_y$. The parameters
used are $t_1=0.18$~eV, $t_2=0.17$~eV, $t_3=0.08$~eV, $t_4=0.02$~eV,
$\mu'-\mu=0.1$~eV, $\tilde{J}_3=-0.1$~eV and $g=0.02$~eV.
\label{phase}}
\end{figure}

Our calculation shows that the spin-orbit coupling lifts the
degeneracy in favor of different pairing state depending on the
relative strength of the pairing interactions, $ \tilde{J}_{2} $ and 
$ \tilde{J}_3 $. In Fig.1 we present the phase diagram of spin-orbit
coupling $ \lambda $ versus the ratio $ \tilde{J}_{2} / \tilde{J}_3
$. Among the three states occurring in the phase diagram we find also
the time reversal symmetry breaking state $ \hat{{\bf z}} (k_x \pm i
k_y) $ which is most stable, if the pairing interaction is dominant in
the $ d_{xy} $-orbital. The other two phase corresponding the
irreducible representations $ A_{1u} $ (phase II) and $ A_{2u} $
(phase I) in Table I, require either substantially stronger pairing
interaction in the bands of the orbitals $ d_{yz} $ and $ d_{zx} $ or
rather weak spin-orbit coupling. The values of $ \lambda $ of the 
Ru$^{4+} $-ion in the literature are of the order of 0.07 eV. 
In view of the effect the spin-orbit coupling has on the quasiparticle
spectrum we may argue that its strength to be moderate, since the 
band structure calculations ignoring spin-orbit coupling are in
reasonable agreement with the de Haas-van Alphen data
\cite{OGUCHI,SINGH,DHVA}. 
There are also experimental indications which suggest that
superconductivity in Sr$_2$RuO$_4$ is carried mainly by the $ d_{xy}
$-band \cite{FORGAN}. Clearly also the electronic density of states of
this band makes up $ 43 \% $ of the whole system. (In the choice of
parameters used in this calculation we have taken the ratios of
density of states into account.) The role of the interband pair
scattering in Eq.(\ref{interband}) is to distinguish between
the $A_{1u} $- and $A_{2u} $-state which would be degenerate
for $ g=0 $. 

Our approach is based on the assumption that spin-orbit coupling is
not too strong so that we can deal with it essentially by
changing only the character of the quasiparticles. This procedure is
complicated due to the itinerant character of the electrons.
In our discussion we have not taken into account the modification of
the spin dependent 
matrix elements of the interaction. For this purpose we would need more
detailed knowledge  
about the pairing mechanism. Nevertheless, for the mechanism based on
the extended Hund's rule coupling we can test the effect of
spin orbit coupling. Calculating the free-quasiparticle spin 
susceptibilities including spin-orbit coupling shows that the
anisotropy in the spin interaction in Eq.(\ref{exchange}) is only of
the order of $ 1 \% $ in the range of $ \lambda $ important for,
considerably smaller than the above effect. 

Calculating the static spin susceptibility for all bands we obtain within
our model a very similar form as calculated from LDA \cite{SF1} and
measured recently by neutron scattering \cite{BRADEN}. Both theory and 
experiment agree well in the aspect that the strongest spin
correlation is not ferromagnetic, $ q= 0 $, but has a finite wave
vector associated with the rather strong nesting feature of the nearly 
one-dimensional bands of the $ d_{yz} $- and $ d_{zx} $-orbitals which
in this way dominate the $ q $-dependence of the susceptibility. 
The large density of states, however, leads to the strongest
contribution to the overall susceptibility by the $ d_{xy} $-band.
In calculating coupling strengths $ J_{2} $ and $ J_3 $ we find that 
their ratio is of the order 1.7 in favor of $ J_{2} $ which still
would lead to the chiral superconducting state in the phase diagram. 
However, also the spin independent interactions have
to be taken into account, which we cannot estimate at present. The main
conclusion of our study is that spin-orbit coupling 
plays an important role in stabilizing the time reversal symmetry
breaking p-wave pairing state in Sr$_2$RuO$_4$ and is strongly
connected with the orbital structure of the electron bands.  

We would like to thank Y. Maeno, Y. Okuno, M. Ogata and T.M. Rice for
helpful discussions. K.K.N. acknowledges support from Japan Society 
for the Promotion of Science. This work was also supported by the
Japanese Ministry of Education, Science, Culture and Sports.


\begin{thebibliography}{99}
\bibitem{MAENO1} Y. Maeno, H. Hashimoto, K. Yoshida, S. Nishizaki, T.
Fujita, J.G. Bednorz and F. Lichtenberg, Nature {\bf 372} (1994) 532.
\bibitem{MACKENZIE} A.P. Mackenzie, R.K.W. Haselwimmer, A.W. Tyler,
G.G. Lonzarich, Y. Mori, S. Nishizaki and Y. Maeno,
Phys. Rev. Lett. {\bf 80} (1998) 161.
\bibitem{RICE} T.~M.~Rice and M.~Sigrist, J. Phys.: Condens. Matter
{\bf 7} (1995) L643. 
\bibitem{BASK} G. Baskaran, Physica B {\bf 223 \&224} (1996) 490. 
\bibitem{LUKE} G.M. Luke, Y. Fudamoto, K.M. Kojima, M.I. Larkin,
J. Merrin, B. Nachumi, Y.J. Uemura, Y. Maeno, Z.Q. Mao, Y. Mori,
H. Nakamura and M. Sigrist, Nature {\bf 394} (1998) 558.
\bibitem{ISHIDA}  K. Ishida, H. Mukuda, Y. Kitaoka, K. Asayama,
Z.Q. Mao, Y. Mori and Y. Maeno, Nature {\bf 396} (1998) 658.
\bibitem{SF1} I.I. Mazin and D.J. Singh, Phys. Rev. Lett. {\bf 82}
(1999) 4324.
\bibitem{HERAKLION} M. Sigrist, D. Agterberg, A. Furusaki, C. Honerkamp,
K.K. Ng, T.M. Rice and M.E. Zhitomirsky, Physica {\bf C 317-318} (1999) 134. 
\bibitem{OGUCHI} T. Oguchi, Phys. Rev. {\bf B51}, 1385 (1995).
\bibitem{SINGH} I.I. Mazin and D.J. Singh, Phys. Rev. {\bf B56} (1997) 2556.
\bibitem{DHVA}  A.P. Mackenzie, S.R. Julian, A.J. Diver, G.G.
Lonzarich, Y. Maeno, S. Nishizaki and T. Fujita, Phys. Rev. Lett. {\bf
76} (1996) 3786.
\bibitem{ODS} D. Agterberg, T.M. Rice and M. Sigrist,
Phys. Rev. Lett. {\bf 78} (1997) 3374.
\bibitem{XIANG} C. Noce and T. Xiang, Physica C {\bf 282-287} (1997) 1713.
\bibitem{FORGAN} T.M. Risemann et al., Nature {\bf 396} (1998) 242.
\bibitem{BRADEN} Y. Sidis, M. Braden, P. Bourges, B. Hennion,
S. NishiZaki, Y. Maeno and Y. Mori, Phys. Rev. Lett. {\bf 83} (1999) 3320.
\bibitem{PRESSURE} K. Yoshida, F. Nakamura, T. Goko, T. Fujita,
Y. Maeno, Y. Mori and S. NishiZaki, Phys. Rev. {\bf B 58} (1998) 8515.
\bibitem{MIYAKE} K. Miyake and O. Narikiyo, Phys. Rev. Lett. {\bf 83}
(1999) 1423. 

\end{thebibliography}
\end{document}